\documentclass[letterpaper,10pt,conference]{templates/ieeeconf}

\IEEEoverridecommandlockouts

\usepackage{amsmath} 

\usepackage{graphicx} 
\usepackage{tabularx} 
\usepackage{multirow} 
\usepackage{balance} 
\usepackage{cite} 

\newcolumntype{C}{>{\centering\arraybackslash}X}
\newcolumntype{L}{>{\raggedright\arraybackslash}X}
\newcolumntype{R}{>{\raggedleft\arraybackslash}X}

\title{\LARGE\bf
    Public Evaluation on Potential Social Impacts of Fully Autonomous\\
    Cybernetic Avatars for Physical Support in Daily-Life Environments:\\
    Large-Scale Demonstration and Survey at Avatar Land
}

\author{
    Lotfi El Hafi$^{1, 2, *}$,
    Kazuma Onishi$^{1}$,
    Shoichi Hasegawa$^{1}$,
    Akira Oyama$^{1}$,
    Tomochika Ishikawa$^{1}$,\\
    Masashi Osada$^{1}$,
    Carl Tornberg$^{1}$,
    Ryoma Kado$^{1}$,
    Kento Murata$^{1}$,
    Saki Hashimoto$^{1}$,\\
    Sebastian Carrera Villalobos$^{2}$,
    Akira Taniguchi$^{1}$,
    Gustavo Alfonso Garcia Ricardez$^{1, 2}$,\\
    Yoshinobu Hagiwara$^{1, 3}$,
    Tatsuya Aoki$^{4}$,
    Kensuke Iwata$^{4}$,
    Takato Horii$^{4}$,
    Yukiko Horikawa$^{5}$,\\
    Takahiro Miyashita$^{5}$,
    Tadahiro Taniguchi$^{1, 6}$,
    and Hiroshi Ishiguro$^{4, 5}$
    \thanks{
        This work was supported by the Japan Science and Technology Agency~(JST), Moonshot Research~\& Development Program, Grant Number JPMJMS2011, and by the Japan Society for the Promotion of Science~(JSPS), KAKENHI Grants-in-Aid for Scientific Research, Grant Numbers JP22K17981, JP23K16975, JP22K17982, and JP22K12212.
    }
    \thanks{
        $^{1}$Lotfi El Hafi, Kazuma Onishi, Shoichi Hasegawa, Akira Oyama, Tomochika Ishikawa, Masashi Osada, Carl Tornberg, Ryoma Kado, Kento Murata, Saki Hashimoto, Akira Taniguchi, Gustavo Alfonso Garcia Ricardez, Yoshinobu Hagiwara, and Tadahiro Taniguchi are with Ritsumeikan University;
        1-1-1 Noji-Higashi, Kusatsu, Shiga 525-8577, Japan. 
        {\tt\small \{lotfi.elhafi, onishi.kazuma, hasegawa.shoichi, oyama.akira, ishikawa.tomochika, osada.masashi, tornberg.carl, kado.ryoma, murata.kento, hashimoto.saki, a.taniguchi, garcia-g, yhagiwara, taniguchi\}@em.ci.ritsumei.ac.jp}
    }
    \thanks{
        $^{2}$Lotfi El Hafi, Sebastian Carrera Villalobos, and Gustavo Alfonso Garcia Ricardez are with Coarobo GK;
        2-2-2 Hikaridai, Seika, Soraku, Kyoto 619-0237, Japan. 
        {\tt\small \{lotfi, sebastian, tavo\}@coarobo.com}
    }
    \thanks{
        $^{3}$Yoshinobu Hagiwara is with Soka University;
        1-236 Tangi, Hachioji, Tokyo 192-8577, Japan. 
        {\tt\small hagiwara@soka.ac.jp}
    }
    \thanks{
        $^{4}$Tatsuya Aoki, Kensuke Iwata, Takato Horii, and Hiroshi Ishiguro are with The University of Osaka;
        1-3 Machikaneyama, Toyonaka, Osaka 560-8531, Japan. 
        {\tt\small \{t.aoki@rlg., k.iwata@rlg., takato@, ishiguro@\}sys.es.osaka-u.ac.jp}
    }
    \thanks{
        $^{5}$Yukiko Horikawa, Takahiro Miyashita, and Hiroshi Ishiguro are with Advanced Telecommunications Research Institute International (ATR);
        2-2-2 Hikaridai, Seika, Soraku, Kyoto 619-0288, Japan. 
        {\tt\small \{horikawa, miyasita, ishiguro\}@atr.jp}
    }
    \thanks{
        $^{6}$Tadahiro Taniguchi is with Kyoto University;
        Yoshida-Honmachi, Sakyo, Kyoto 606-8501, Japan. 
        {\tt\small taniguchi@i.kyoto-u.ac.jp}
    }
    \thanks{
        $^{*}$Corresponding author.
    }
}

\begin{document}


\maketitle
\thispagestyle{empty}
\pagestyle{empty}


\begin{abstract}
    Cybernetic avatars (CAs) are key components of an avatar-symbiotic society, enabling individuals to overcome physical limitations through virtual agents and robotic assistants.
    While semi-autonomous CAs intermittently require human teleoperation and supervision, the deployment of fully autonomous CAs remains a challenge.
    This study evaluates public perception and potential social impacts of fully autonomous CAs for physical support in daily life.
    To this end, we conducted a large-scale demonstration and survey during Avatar Land, a 19-day public event in Osaka, Japan, where fully autonomous robotic CAs, alongside semi-autonomous CAs, performed daily object retrieval tasks.
    Specifically, we analyzed responses from 2,285 visitors who engaged with various CAs, including a subset of 333 participants who interacted with fully autonomous CAs and shared their perceptions and concerns through a survey questionnaire.
    The survey results indicate interest in CAs for physical support in daily life and at work.
    However, concerns were raised regarding task execution reliability.
    In contrast, cost and human-like interaction were not dominant concerns.
    Project page: https://lotfielhafi.github.io/FACA-Survey/.
\end{abstract}


\begin{figure}[t]
    \centering
    \includegraphics[width=0.93\linewidth]{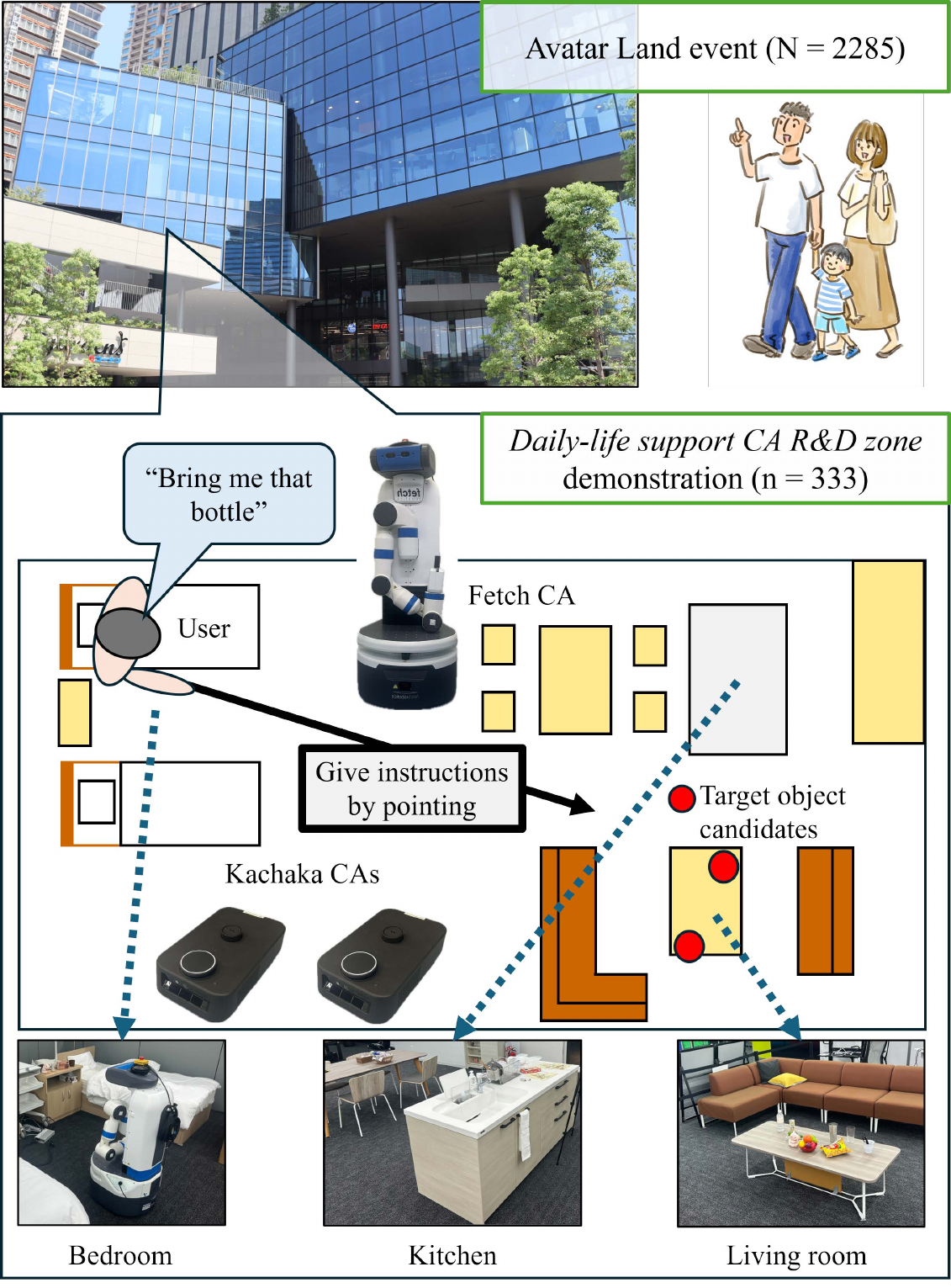}
    \caption{
        Overview of the \textit{daily-life support CA R\&D zone} demonstration at Avatar Land, where visitors interacted with fully autonomous cybernetic avatars (CAs) in a replicated home environment.
    }
    \label{fig:research_overview}
\end{figure}


\section{Introduction}

As birth rates decline and the population ages in developed societies, an increasing burden is expected to be placed on the working population to care for younger and elderly individuals while maintaining economic productivity and social welfare.
In this context, cybernetic avatars (CAs) have been envisioned as a key component of an avatar-symbiotic society~\cite{ishiguro_realisation_2021}, aiming to free individuals from physical constraints by expanding their capabilities through teleoperated, semi-autonomous virtual agents and robotic assistants.
While the application of CAs as conversational agents has been actively researched and demonstrated~\cite{sakai_simultaneous_2024, horikawa_cybernetic_2023, fu_dual_2023}, broader adoption is yet to be achieved, and further public assessment is required to demonstrate their necessity.
In particular, most CA-related research has focused on semi-autonomous robots reliant on human teleoperation.
However, the feasibility and social implications of fully autonomous robotic CAs and their physical interaction capabilities remain underexplored.

In this regard, various studies have investigated using service robots to provide physical support in daily life and work settings.
Recent advancements in foundation models for robotics~\cite{ghosh_octo_2024, black_pi0_2024, oneill_open_2024} aim to enable robots to acquire general-purpose physical skills by learning from large-scale datasets collected in home and office environments.
However, achieving full substitution of human tasks remains a significant challenge.
This is further highlighted in robotics competitions that turn service tasks, including object manipulation and instructions in natural language, into competitive benchmarks~\cite{wada_new_2017, contreras_towards_2022}.

Many robots that assist humans require high degrees of freedom to operate effectively across diverse environments and objects.
While recent mobile manipulators and humanoid robots are capable of dexterous full-body movements, they must also dynamically perceive their surroundings at all times to ensure contextual awareness, shared understanding, and safe operations, which remain challenging in daily environments alongside humans.
Therefore, gathering public data for validating future research and development directions toward fully autonomous CAs capable of physical interaction is essential, which requires publicly demonstrating prototypes to a representative sample of non-expert users.

To this end, we conducted a large-scale demonstration and survey during Avatar Land\footnote{https://avatar-ss-land.iroobo.jp/}, a month-long event in a densely populated area of Osaka, Japan, where fully autonomous robotic CAs, alongside semi-autonomous robotic CAs, performed daily object retrieval tasks, as illustrated by Fig.~\ref{fig:research_overview}.
This demonstration was open to the public, and all visitors were invited to complete a survey questionnaire assessing their experiences with various CAs, including fully autonomous robotic systems.
Following Avatar Land, we conducted a comprehensive analysis of the responses from 2,285 visitors, evaluating public perceptions, potential demand, and societal concerns regarding fully autonomous CAs.
Hence, instead of focusing on quantitative metrics, our approach centered on assessing subjective public perception on a large scale.

Our contributions are threefold:
\begin{itemize}
    \item We developed and deployed a fully autonomous CA demonstration, distinguishing it from the semi-autonomous approaches commonly explored with CAs.
    \item We conducted a large-scale public survey to assess the potential social impacts of fully autonomous CAs providing physical support in daily life.
    \item We analyzed the survey data to derive insights into public perception toward the future deployment of fully autonomous CAs.
\end{itemize}


\begin{figure*}[t]
    \centering
    \includegraphics[width=1.0\linewidth]{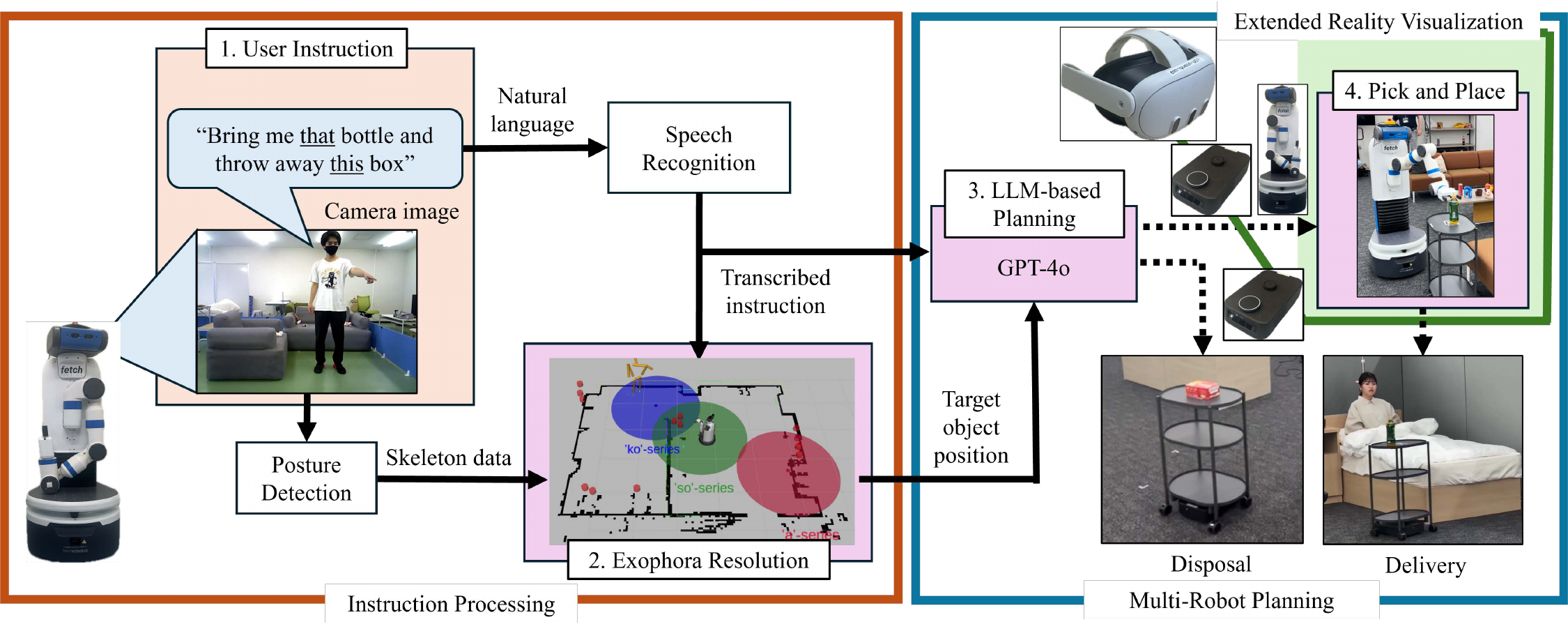}
    \caption{
        Overview of the implementation of the demonstrated system.
        The user provides instructions to the robot by pointing at the target object.
        The system performs exophora resolution to identify the intended object and autonomously plans task execution among multiple robotic CAs.
    }
    \label{fig:implementation_overview}
\end{figure*}


\section{Survey Methodology}

Avatar Land was held at Grand Green Osaka, a newly opened commercial and public facility located a 7-minute walk from Osaka Station, one of Japan’s largest transportation hubs.
The event ran from September 10 to 29, 2024, with daily opening hours from 10:00 to 16:00.
The location and open nature of the event allowed for the gathering of a general audience, including families with young children and elderly individuals.
Visitors could participate in 11 distinct interactive demonstrations illustrating various aspects of CAs' potential applications, at their own pace, with staff members providing guidance and explanations, and inviting them to fill a survey questionnaire at the end of their visit.
The demonstrations were named as follows: \textit{computer graphics (CG) CA zone}, \textit{CA receptionist zone}, \textit{communication training}, \textit{caregiving support CA zone}, \textit{CA teleoperation zone}, \textit{daily-life support CA R\&D zone}, \textit{single operator controlling 15 CAs}, \textit{facility guidance from teleoperated mobile CA}, \textit{guide service from various CA cooperation}, \textit{multi-language service for international cooperation}, and \textit{chat service from paired CAs}.

Filling out the survey questionnaire was voluntary, and responses were collected from 2,285 participants who chose to answer ($N = 2285$).
The participants who volunteered were asked the following multiple-choice questions:
\begin{itemize}
    \item Q1: ``Which CA demonstration did you participate in?''
    \item Q2: ``Would you use the demonstrated CAs in your daily life?''
    \item Q3: ``In what situations would you use the demonstrated CAs?''
    \item Q4: ``Why do you not want to use the demonstrated CAs?''
\end{itemize}
Q1 was asked once to all survey volunteers, while Q2–Q4 were asked for each demonstration they participated in.
The detailed possible answers are provided in Table~\ref{tab:consolidated_results} and Fig.~\ref{fig:consolidated_results}.
All data were anonymized, and visitors were informed that their data could be used for research purposes by the ethical guidelines of the Japan Science and Technology Agency~(JST).
Demographic data were also collected, but their analysis is outside the scope of this study.
The total number of visitors is unknown as the demonstration site was deliberately open, without controlled entry points to encourage spontaneous interaction with the CAs.
However, the number of visitors who did not respond to the survey questionnaire was likely an order of magnitude higher than those who did.

The demonstration described in this study, \textit{daily-life support CA R\&D zone}, was proposed by Group 4\footnote{https://avatar-ss.org/en/members/index.html} within the JST Moonshot Research~\& Development Program project Avatar-Symbiotic Society~\cite{ishiguro_realisation_2021}.
Importantly, the \textit{daily-life support CA R\&D zone} was the only demonstration showing fully autonomous robotic CAs that were not teleoperated in any form, making it unique in its emphasis on autonomous physical support in daily life.
To this aim, it took place in a reproduced home environment that included a bedroom, kitchen, living room, and medical home care equipment, as described in Fig.~\ref{fig:research_overview}.
The demonstrated system consisted of three fully autonomous robotic CAs: a Fetch Robotics Fetch\footnote{http://docs.fetchrobotics.com/} mobile manipulator and two Preferred Robotics Kachaka\footnote{https://kachaka.life/home/} shelf-carrying robots.

Participants were invited to observe and interact with the system in a scenario where a user in bed would point to a target object and instruct the robot to retrieve it.
The system performed exophora resolution to distinguish the target from similar objects using multimodal contextual information gathered from active exploration, ensuring accurate task execution.
The robotic then CAs collaborated to execute the task autonomously, while visitors could monitor the process through a Meta Quest 3\footnote{https://www.meta.com/quest/quest-3/} extended reality (XR) headset, which provided real-time visualization of the robotic CAs' perception and execution processes.
Additional TV displays were installed to share real-time visualization outside the XR headset and encourage visitor engagement.

The \textit{daily-life support CA R\&D zone} demonstration was designed as a real-world use case aligned with the envisioned future of an avatar-symbiotic society, showcasing how fully autonomous CAs could provide physical support in daily-life environments with:
\begin{itemize}
    \item Multi-robot collaboration that allows a single user to interact with multiple CAs.
    \item Full autonomy that eliminates human supervision or teleoperation of the CAs.
    \item Intuitive interaction that relies on pointing gestures and natural language instructions to the CAs.
    \item Real-time visualization that facilitates shared understanding in XR between the user and CAs.
\end{itemize}
From the 2,285 survey questionnaires collected during Avatar Land, 333 visitors reported participating in \textit{daily-life support CA R\&D zone} demonstration ($n = 333$).


\begin{table}[t]
    \caption{
        Consolidated Survey Questionnaire Results
    }
    \begin{center}
        \begin{tabularx}{1.0\linewidth}{|L|r|r|}
            \hline
            \multicolumn{3}{|c|}{\textbf{Demonstration Participation ($\mathbf{N = 2285}$)}} \\
            \hline
            \multicolumn{3}{|c|}{``Q1: Which CA demonstration did you participate in?''} \\
            \hline
            Responses & Rate [\%] & Sample \\
            \hline
            CG CA zone & 56.4 & 1289 \\
            CA receptionist zone & 47.9 & 1095 \\
            Communication training & 31.1 & 711 \\
            Caregiving support CA zone & 39.7 & 907 \\
            CA teleoperation zone & 16.9 & 386 \\
            \textbf{Daily-life support CA R\&D zone} & \textbf{14.7} & $\mathbf{n = 333}$ \\
            Single operator controlling 15 CAs& 25.9 & 591 \\
            Facility guidance from teleop. mobile CA & 31.1 & 710 \\
            Guide service from various CA coop. & 12.4 & 283 \\
            Multi-language service for int'l coop.  & 9.1 & 207 \\
            Chat service from paired CAs & 7.2 & 164 \\
            No answer & 2.1 & 47 \\
            \hline
            \hline
            \multicolumn{3}{|c|}{\textbf{Usage Likelihood ($\mathbf{n = 333}$)}} \\
            \hline
            \multicolumn{3}{|c|}{``Q2: Would you use the demonstrated CAs in your daily life?''} \\
            \hline
            Responses & Rate [\%] & Sample \\
            \hline
            \textbf{Very likely to use} & \textbf{39.3} & $\mathbf{n^+_1 = 131}$ \\
            \textbf{Use if conditions are right} & \textbf{35.4} & $\mathbf{n^+_2 = 118}$ \\
            Cannot tell if want to use or not & 20.8 & 69 \\
            \textbf{Do not want to use} & \textbf{2.4} & $\mathbf{n^- = 8}$ \\
            No answer & 2.1 & 7 \\
            \hline
            \hline
            \multicolumn{3}{|c|}{\textbf{Usage Scenarios ($\mathbf{n^+_1 + n^+_2 = 249}$)}} \\
            \hline
            \multicolumn{3}{|c|}{``Q3: In what situations would you use the demonstrated CAs?''} \\
            \hline
            Responses & Rate [\%] & Sample \\
            \hline
            Work & 32.5 & 81 \\
            Education and studying & 17.3 & 43 \\
            Hobby and leisure & 15.7 & 39 \\
            Conversation and communication & 15.3 & 38 \\
            Daily life & 47.8 & 119 \\
            Others & 4.4 & 11 \\
            No answer & 3.2 & 8 \\
            \hline
            \hline
            \multicolumn{3}{|c|}{\textbf{Usage Aversions ($\mathbf{n^- = 8}$)}} \\
            \hline
            \multicolumn{3}{|c|}{``Q4: Why do you not want to use the demonstrated CAs?''} \\
            \hline
            Responses & Rate [\%] & Sample \\
            \hline
            Could not use well & 37.5 & 3 \\
            Seemed fragile & 12.5 & 1 \\
            Seemed expensive & 12.5 & 1 \\
            No space for using or storing & 0.0 & 0 \\
            Prefer human face-to-face interaction & 25.0 & 2 \\
            Others & 25.0 & 2 \\
            No answer & 25.0 & 2 \\
            \hline
            \hline
            \multicolumn{3}{|c|}{\textbf{Participant Comments}} \\
            \hline
            \multicolumn{3}{|p{0.94\linewidth}|}{C1: ``I felt that life-support avatars are easy to imagine being used in various situations and can be utilized in many ways. However, I could not clearly envision their applications in communication, so I am looking forward to seeing how they will be used in the future.''} \\
            \multicolumn{3}{|p{0.94\linewidth}|}{C2: ``I would love for them to be useful in households, especially for tasks like fetching items and tidying up.''} \\
            \multicolumn{3}{|p{0.94\linewidth}|}{C3: ``It was fascinating to see the perspective of an autonomous avatar (robot). I was surprised at how much calculation goes into even picking up a single object.''} \\
            \multicolumn{3}{|p{0.94\linewidth}|}{C4: ``I believe that if remote-support robots become more practical, they will greatly expand the possibilities in our daily lives.''} \\
            \multicolumn{3}{|p{0.94\linewidth}|}{C5: ``I have high expectations for life-support and security robots (avatars). I think they could contribute to community safety.''} \\
            \multicolumn{3}{|p{0.94\linewidth}|}{C6: ``Life-support assistance when caregiving is needed, as well as customer service in public facilities.''} \\
            \multicolumn{3}{|p{0.94\linewidth}|}{C7: ``When I used a caregiving avatar to carry a plastic bottle, I wished it could also open the bottle cap.''} \\
            \hline
        \end{tabularx}
    \end{center}
    \label{tab:consolidated_results}
\end{table}


\section{System Overview}

This section details the flow and technical implementation of the \textit{daily-life support CA R\&D zone} demonstration, as described by Fig.~\ref{fig:implementation_overview}.
The system enables fully autonomous CAs to process user instructions, plan multi-robot actions, and execute object retrieval tasks.
The entire demonstration was implemented using a containerized software development environment~(SDE)~\cite{el_hafi_software_2022}, with all communication between CAs handled via ROS~\cite{quigley_ros:_2009}.

User speech was transcribed into text using Whisper~\cite{radford_robust_2023}, while the user posture was extracted from RGB-D images captured by Fetch CA's head camera using MediaPipe~\cite{lugaresi_mediapipe_2019}.
These inputs were processed by an exophora resolution model~\cite{oyama_exophora_2023}, enabling the system to determine the target object based on pointing gestures and natural-language instructions, and contextual information.
The identified object location and user command were input into a large language model (LLM), GPT-4o~\cite{brown_language_2020}, for multi-robot task planning.

The task execution sequence involved Fetch and one Kachaka robot CAs retrieving the specified object, while the second Kachaka CA handled object disposal.
The entire process could be monitored in real-time using a Meta Quest 3 XR headset, providing users with an interactive visualization of robot CAs' perception and execution.


\subsection{Instruction Processing}

User instructions were processed using an exophora resolution model~\cite{oyama_exophora_2023}, which resolves ambiguous linguistic commands, such as those including pronouns or demonstratives, through contextual information.
Exophora information, in contrast to endophora information, refers here to information not included in the user's instruction itself but inferred based on contextual information.
The probability $p$ of the estimated target object location was computed as the product of three probabilities: $p_1$ related to the user's pointing gesture direction, $p_2$ related to the demonstrative word used in the instructions, and $p_3$ related to the object's category distribution in the environment.

In this demonstration, $p_1$ was determined using pose skeleton data, specifically the user's eyes and wrists, extracted from images captured by Fetch CA's head camera and processed with MediaPipe~\cite{lugaresi_mediapipe_2019}.
$p_2$ was assigned based on the detected demonstrative word in the user’s instruction (\textit{e.g.}, ``this'' or ``that'').
Ideally, $p_3$ should be derived from active exploration methods~\cite{ishikawa_active_2023}, to automatically acquire contextual information and facilitate the introduction of robotic CAs in new environments without user intervention.

However, to ensure the robustness of exophora resolution during the \textit{daily-life support CA R\&D zone} demonstration, we preloaded a user's pose skeleton and transcribed instruction data prepared offline, and manually recorded object locations as the model inputs.
In addition, instead of demonstrating active exploration with visitors, which is time-consuming by nature, a pre-recorded video of the exploration performed on the site of Avatar Land was shown on TV displays.


\subsection{Object Manipulation}

After moving near the target object inferred by the exophora resolution model, the Fetch CA picks up the object and places it on a Kachaka CA for delivery to the user, as the \textit{daily-life support CA R\&D zone} demonstration was designed to emphasize multi-CA collaboration.
Although the position of the object is assumed to be known, in reality, there may be errors in self-position estimation, or the object position may have changed since the exophora resolution.
Therefore, additional object detection was conducted with Detic~\cite{zhou_detecting_2022} before grasping the target object for confirmation.
The detected results were then combined with depth information to estimate the position of the target object.
The position for placing the object on the Kachaka CA also needed to be determined similarly, so AR markers were attached to the shelf to compute the relative positions between the two CAs.
MoveIt's RRT-Connect\footnote{https://ompl.kavrakilab.org/} was used for motion planning to pick and place the target object.

This conventional manipulation approach was also augmented with an experimental grasping method using infrared light-based proximity sensors~\cite{suzuki_grasping_2022}.
We attached these sensors to the Fetch CA's fingertip and used them to control hand positioning and the grasping approach motion by inferring the target object's reflectance using a multimodal LLM, which is a unique property of each material~\cite{garcia_ricardez_reflectance_2023}.
However, this experimental proximity sensing-based grasping system was not demonstrated throughout the entire Avatar Land event, as we reverted to the conventional approach due to wiring limitations between the proximity sensors and the Fetch CA, which could have caused accidents.


\subsection{Multi-Robot Planning}

As the \textit{daily-life support CA R\&D zone} demonstration was designed to emphasize multi-CA collaboration, task allocation between Fetch and the two Kachaka robotic CAs was performed with an LLM-based approach~\cite{hasegawa_reducing_2025} using GPT-4o.
In this approach, appropriate action sequences were generated for each CA by providing GPT-4o with the capabilities of each robot and an example of task planning.
However, to ensure reliable task planning during the \textit{daily-life support CA R\&D zone} demonstration, we precomputed task allocations using a set of predefined user instructions.
The execution itself of each task was managed using a FlexBE behavior engine~\cite{schillinger_human-robot_2016} state machine.

The demonstration also required ROS communication between all three robotic CAs.
As Kachaka operates on ROS~2, we wrapped Kachaka’s Python API in a ROS-compatible interface, enabling unified control of all robots inside our containerized SDE~\cite{el_hafi_software_2022} running ROS.


\subsection{Extended Reality Visualization}

For users to better understand the system's operation and monitor task execution of fully autonomous CAs in real time, a visualization interface was developed based on recent findings in human-robot interaction in XR~\cite{el_hafi_teaching_2021, nakamura_multimodal_2022, tornberg_mixed_2024}.
Specifically, a Meta Quest 3 XR headset was used to display real-time sensor data and robot motion plans, allowing users to observe the CAs' self-localization, navigation paths, and planned arm trajectories.
Importantly, the user could leverage the passthrough cameras of the XR headset to adjust the level of virtuality between augmented reality (AR), augmented virtuality (AV), and virtual reality (VR).


\begin{figure}[t]
    \centering
    \includegraphics[width=1.0\linewidth]{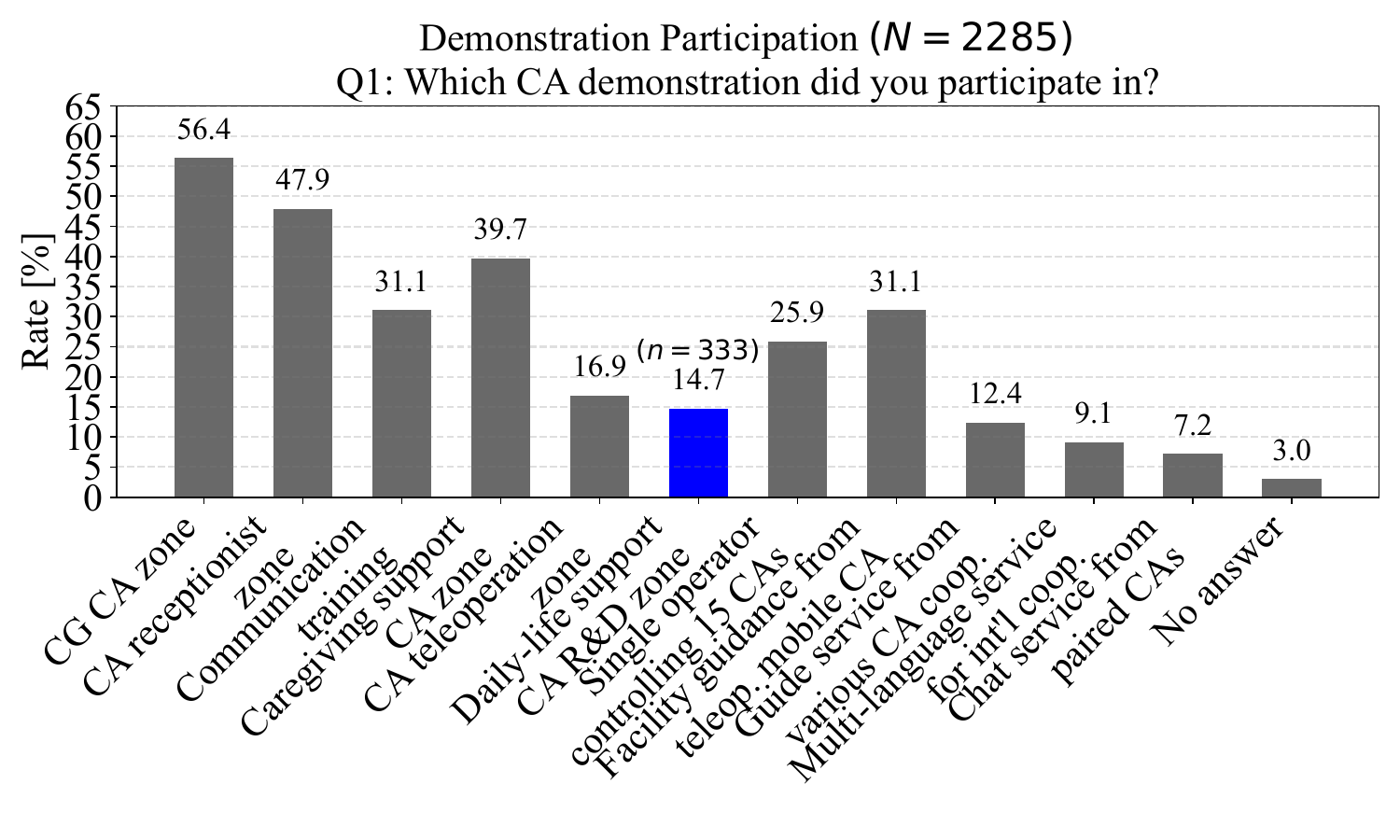}\\
    \includegraphics[width=1.0\linewidth]{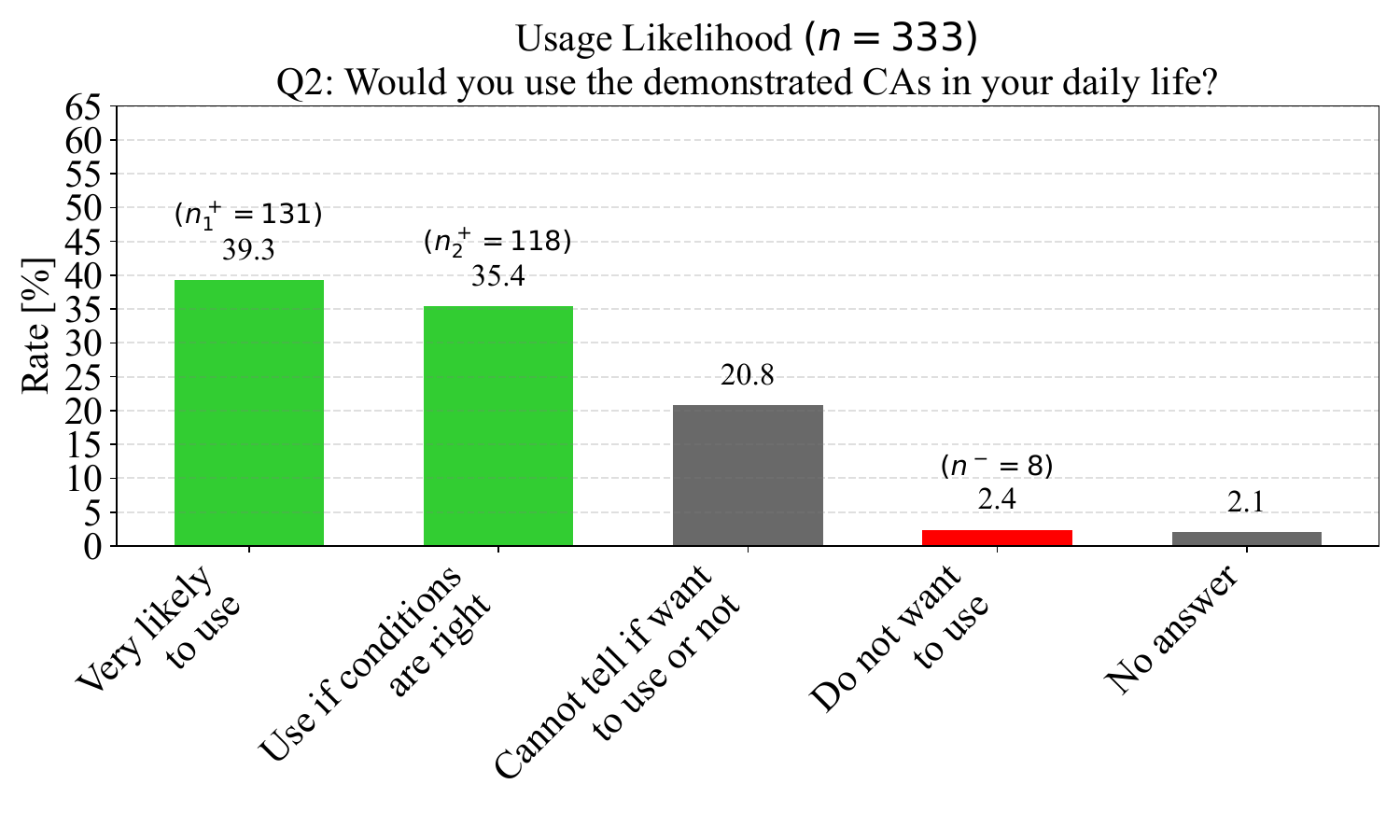}\\
    \includegraphics[width=1.0\linewidth]{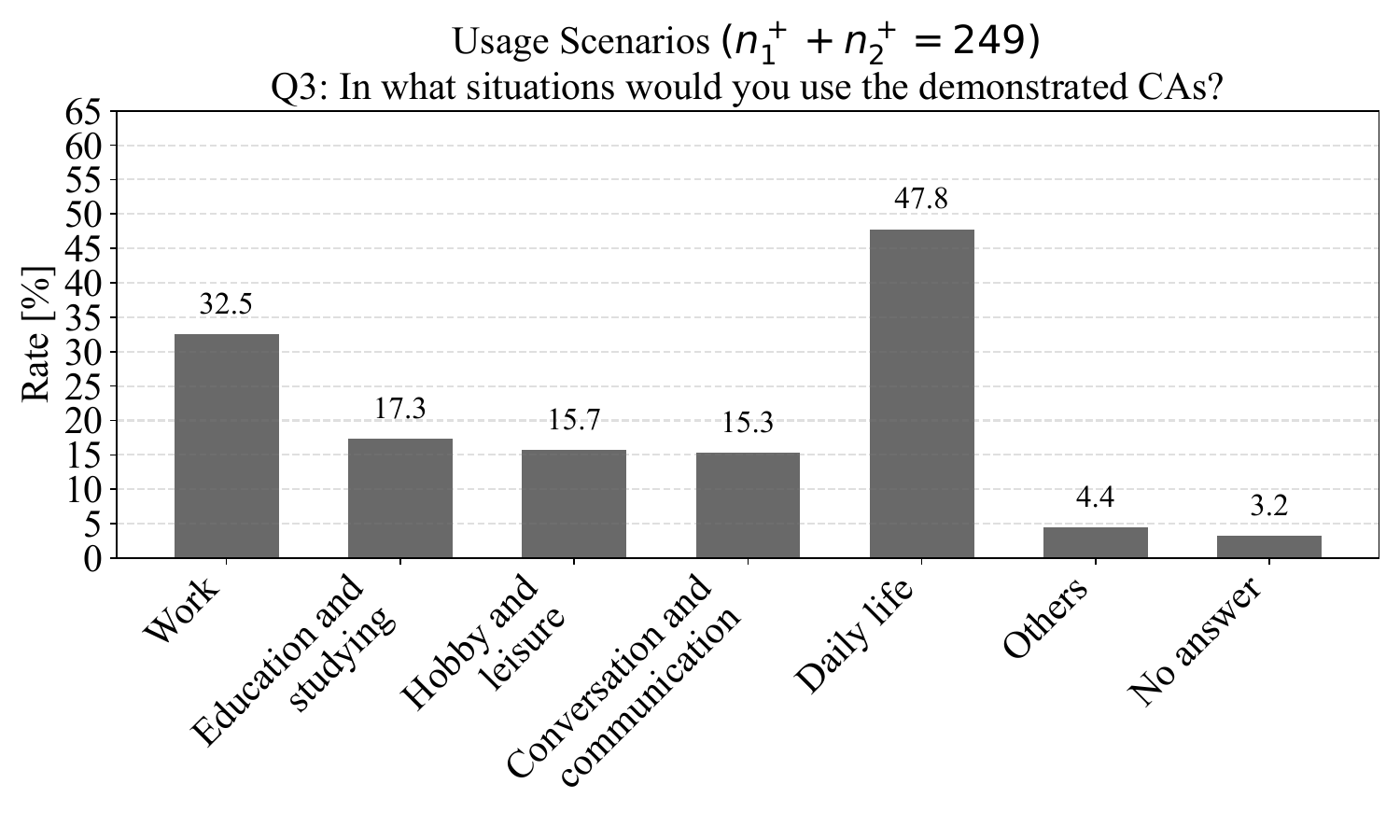}\\
    \includegraphics[width=1.0\linewidth]{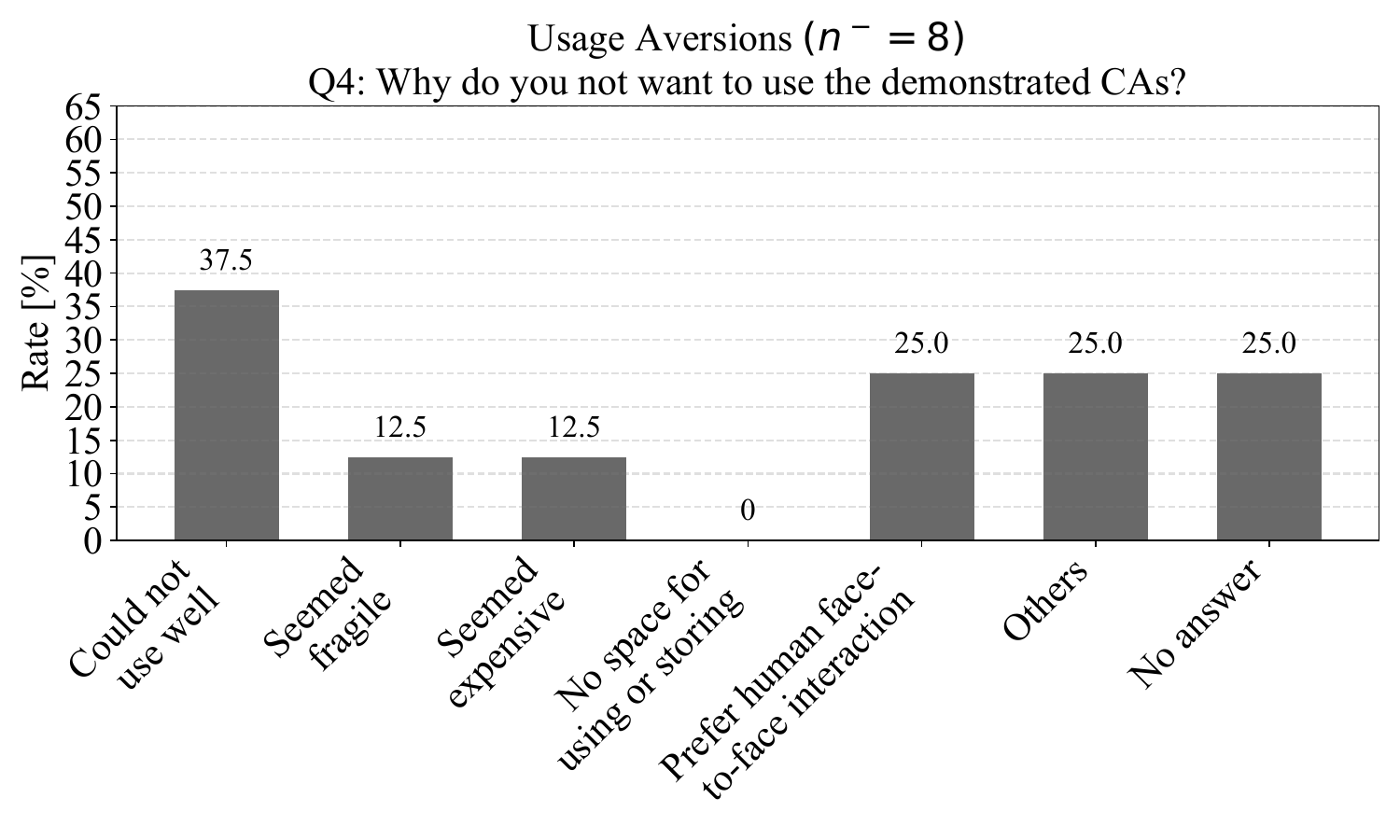}
    \caption{
        Consolidated survey questionnaire results.
    }
    \label{fig:consolidated_results}
\end{figure}


\section{Result Analysis}

The survey results are shown in Table~\ref{tab:consolidated_results} and Fig.~\ref{fig:consolidated_results}.
Fig.~\ref{fig:consolidated_results}(Q1) describes the participation of the 2,285 survey volunteers ($N = 2285$) across all 11 demonstrations, given the question ``Which CA demonstration did you participate in?'' (Q1).
A total of 333 respondents participated in the \textit{daily-life support CA R\&D zone} demonstration described in this study ($n = 333$), ranking 8th out of the 11 demonstrations.

The usage likelihood regarding the demonstrated fully autonomous CAs is shown in Fig.~\ref{fig:consolidated_results}(Q2).
Among those who experienced the \textit{daily-life support CA R\&D zone}, 39.3\% answered ``Very likely to use'' ($n^+_1 = 131$), and 35.4\% responded ``Use if conditions are right'' ($n^+_2 = 118$) to the question ``Would you use the demonstrated CAs in your daily life?'' (Q2).
This implies that 74.7\% of the respondents ($n^+_1 + n^+_2 = 249$) had a positive attitude toward CAs providing daily-life support, indicating that the demonstration generated public interest.
It is important to acknowledge that these positive responses may reflect initial impressions rather than views on long-term usability.
Additionally, the results suggest that aversion to daily-life support CAs is minimal, with only 2.4\% of ``Do not want to use'' responses ($n^- = 8$).

The scenarios in which people would like to use CAs are shown in Fig.~\ref{fig:consolidated_results}(Q3).
To the question ``In what situations would you use the demonstrated CAs?'' (Q3), the most frequent responses were for ``Daily life'' and at ``Work'', with 47.8\% and 32.5\% of responses, respectively.
Since the demonstration focused on physical daily-life assistance, it is notable that the percentage of ``Daily life'' answers did not exceed 50\%.
This suggests a gap in the perceived applicability of CAs for physical support in home environments.
Further investigation is required to understand the factors contributing to this gap.

Fig.~\ref{fig:consolidated_results}(Q4) highlights the reasons why some respondents were hesitant to use CAs.
To the question ``Why do you not want to use the demonstrated CAs?'' (Q4), 37.5\% of respondent mentioned ``Could not use well'' as their primary concern.
From the results of Fig.~\ref{fig:consolidated_results}(Q2) and Fig.~\ref{fig:consolidated_results}(Q4), ensuring a high success rate and stability in demonstrations appears critical to meeting users' conditions for adopting fully autonomous CAs at home.

We also gathered feedback from the free-text section of the survey, as reported at the bottom of Table~\ref{tab:consolidated_results}.
Some respondents, like in comment C7, suggested that adaptable manipulation for daily tasks should be prioritized to enhance perceived utility.
Others expressed hope for the practical application of robotic CAs and noted that their interest in the technology increased after experiencing the demonstration, suggesting that such public engagement events are valuable in helping non-experts envision the future of CAs.
However, free-text comments were too few to draw strong conclusions.


\section{Conclusion}

This study presented a large-scale public demonstration and survey at Avatar Land, evaluating public perception and the potential social impact of fully autonomous CAs in daily-life environments.
Unlike prior research on teleoperated CAs, this study assessed how fully autonomous robotic CAs were perceived when performing physical support tasks without human supervision.
Specifically, we conducted a comprehensive analysis of responses from 2,285 visitors engaging with various CAs, from which a subset of 333 participated in the demonstration of fully autonomous CAs described in this study.
Instead of focusing on quantitative metrics, our approach centered on assessing subjective public perception on a large scale via a survey questionnaire.

The survey results indicated public interest in integrating fully autonomous CAs into daily life and at work.
However, concerns were raised about task execution reliability, with hesitation primarily centered on whether the robots could consistently complete tasks successfully.
In contrast, factors such as cost and human-like interaction were not dominant concerns, although perceptions of cost might evolve with more prolonged user experience.
These findings highlight the need to improve robot CA task performance to increase public adoption toward realizing an avatar-symbiotic society.






\balance


\bibliographystyle{templates/IEEEtran}
\bibliography{references}


\end{document}